 \documentclass[
prl,twocolumn,showpacs, amsmath, amssymb, floatfix,
nofootinbib,wrapfloat byrevtex]{revtex4}

 \usepackage{amsmath,mathrsfs,upgreek,graphicx,epstopdf,placeins,float}
 \usepackage{color}
 \usepackage{booktabs}
 \usepackage{multirow}

 \newcommand{\beq}[1]{\begin{equation}\label{#1}}
 \newcommand{\eeq}{\end{equation}}
 \newcommand{\bea}[1]{\begin{eqnarray}\label{#1}}
 \newcommand{\eea}{\end{eqnarray}}
 \newcommand{\m}{{\rule[1.5pt]{2pt}{0.5pt}}}
 \newcommand{\mm}{{\rule{2pt}{0.5pt}}}

\begin{document}

 \title{Information Missing Puzzle, Where Is Hawking's Error?}
 \author{Ding-fang Zeng}
 \email{dfzeng@bjut.edu.cn}
 \affiliation{Institute of Theoretical Physics, Beijing University of Technology, China, Bejing 100124}
 \begin{abstract}
Matters falling into and consisting of a blackhole can oscillate periodically across instead of accumulate statically on the central point and form singularities there. In quantum language, this oscillation not only resolves central singularities of the blackhole but also blurs its horizon remarkably. This blurring makes the horizon not a zero-thickness geometric surface any more, but an extended physic region whose thickness is comparable with the horizon radius itself. It is our negligence of this fact that leads to the information missing puzzle, and other related question in blackhole physics. Besides the title question and Schwarzschild singularity's resolving, the current work also provides interpretations for the origin of Bekenstein-Hawking entropy and an explicitly unitary formulation of Hawking radiations. 
 \end{abstract}
 \maketitle

\section{Introduction} Steven Hawking walked through his extraordinary life several months ago, leaving us great scientific heritages among which the discovering that blackholes can radiate in such a way that information may miss \cite{hawking1975cmp,hawking1976,wald1975}  attracts the world's attention most deeply.  Since information missing breaks the basic principle of quantum mechanics\footnote{See e.g. \cite{QMshankar}, The quantum state of a close system evolving unitarily --- obeying Schrodinger equation --- is considered as a basic principle of quantum mechanics. By this principle pure state can not become mixed one. The blackhole and its radiation products as a whole can be looked as a close system.}, many physicists consider it a huge conflict between general relativity and quantum mechanics and a strong signal to necessities of the more fundamental theory of quantum gravitations. For this reason, physicists including Hawking himself pay much effort to examine if this is indeed the case or not, see refs.\cite{dfzeng2017,dfzeng2018} for more related works. Basing on general ideas of gauge/gravity duality, most researchers in this area agree that the basic principle of quantum mechanics can not be violated and Hawking must do something wrong somewhere in the calculation or reasonings. However, consensus on where and what he does erroneously that lead to this puzzle is never reached uniformly in the area \cite{mathur2009,polchinski2016}. The purpose of this work is to provide a new answer to this question which has pictures relevant with but different from the string theory fuzzballs \cite{LuninMathur0202,mathur0502} and uses ideas only of standard general relativity and canonic quantum mechanics. 

\begin{figure}[ht]
\begin{center}
\includegraphics[scale=0.8,clip=true,bb=0 40 252 164]{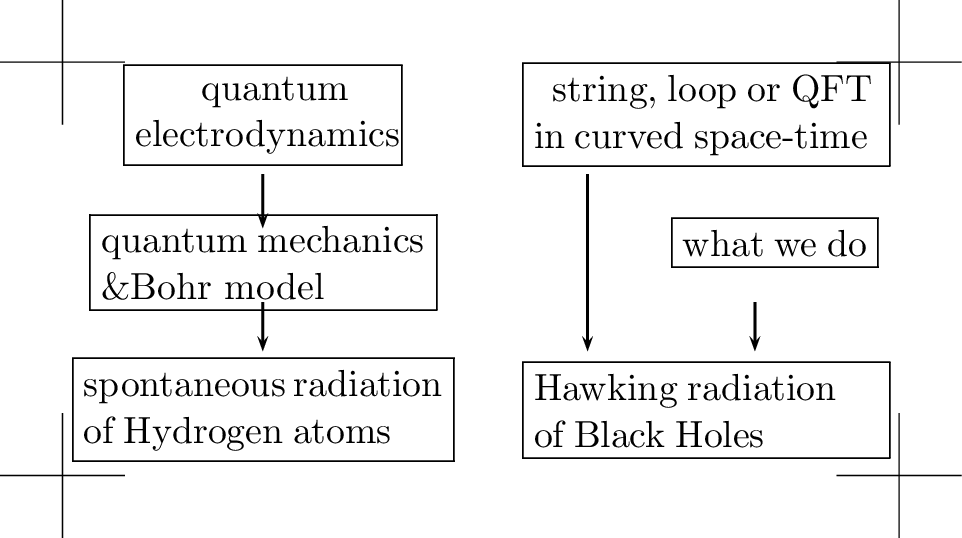}
\end{center}
\caption{What we do in this work between quantum gravitation and Hawking radiations can be looked as a counterpart of what quantum mechanics do between quantum electrodynamics and hydrogen atom spectrums.}
\label{figFramework}
\end{figure}

From the viewpoint of full quantum gravitation theories, FIG.\ref{figFramework}, our answer lies on a level similar to that of quantum mechanics in the working flow of quantum electrodynamics and Hydrogen atom physics. This is a valuable working level rarely explored by previous researchers. Bekenstein and Mukhanov \cite{bekenstein1997} once tried to provide explanations for the blackhole entropy and quantisation at the same level but failed. In their tryings, the microscopic degree of freedom of blackholes is carried by the horizon area elements directly and the hole has discrete mass levels, which leads to line-shape radiation spectrum, thus contradicting with Hawking's radiation spectrum explicitly. While in our works, the blackhole microscopic degrees of freedom is encoded in the radial eigen motion modes of its matter-energy contents and no discrete mass spectrum is required of the hole itself. Due to the same reason, to get area law entropy in our picture is a highly non-trivial work. We provide persuasive evidences following from concrete calculations in the main text that this is possible and reasonable. Our conclusion is that, using standard general relativity and canonical quantum mechanics, we can provide perfectly consistent answers to most of the subtle/puzzling events occurring on both the central point and horizon surface of blackholes. More importantly, our answer is dis/verifiable in the high precision gravitational wave observations.

Our work may not be the final answer to the information missing puzzle. But it is of great possibility that we are on the right way, that is, quantum aspects of gravitation collapse, the central singularity's resolving and the essence of blackhole microscopic state must be considered in this question. On this point, our work agrees well with previous works such as \cite{Stojkovic2008a,Stojkovic2008b,Stojkovic2009,Stojkovic2014,stojkovic2015,Stojkovic2016}

\section{ Where is the error? --- Simple but straight forward answer} The key idea behind Hawking's calculation and reasoning is rather simple \cite{polchinski2016}. Surrounding a blackhole, both the freely falling observer and fixed position ones have their own mode expansion for quantum fields, 
\beq{}
\left\{\begin{array}{l}\displaystyle\phi=\sum_\omega\big(a^\mathrm{fr}_\omega e^{-i\omega u}+a^{\mathrm{fr}\dagger}_\omega e^{i\omega u}\big),~u\sim t-r
\vspace{1mm}\\
\phi=\displaystyle\sum_\omega\big(a^\mathrm{fx}_\omega e^{-i\omega\bar{u}}+a^{\mathrm{fx}\dagger}_\omega e^{i\omega \bar{u}}\big),~\bar{u}=\bar{u}(u)
\end{array}\right.
\eeq
\beq{}
a^\mathrm{fx}_\omega=\sum_k(\alpha_{\omega\kappa}a^\mathrm{fr}_\kappa
+\beta^*_{\omega\kappa}a^{\mathrm{fr}\dagger}_\kappa),~c.c.
\eeq
However, after some pure technique derivations, Hawking finds that the vacuum state identified by freely falling observers are not vacuums of the fixed position ones,
\beq{}
\langle\mathrm{vac}|a^{\mathrm{fr}\dagger}_\omega\!a^\mathrm{fr}_\omega|\mathrm{vac}\rangle=0
\neq\langle\mathrm{vac}|a^{\mathrm{fx}\dagger}_\omega\!a^\mathrm{fx}_\omega|\mathrm{vac}\rangle
=\frac{1}{e^\frac{\hbar\omega}{kT}\!\pm\!1}
\eeq
That is, the latters see particles radiating from the blackhole horizon and the thermal feature of these particles implies that pure states describing the pre-blackhole contents evolve into mixed ones of the radiation products so that the unitarity principle of quantum mechanics is violated during this process. 

Directive as it is, Hawking's train of thought contains a key bug. It focuses on particles being measured outside the horizon but neglects the fact that blackhole has inner structure and microstate, which is just the goal he tries to show against but provides evidences for in his original work on this question \cite{hawking1975cmp}. Just as we pointed out in ref.\cite{dfzeng2018},  thermal features of the radiation products can be derived out from averages on the initial and summations on the final state of blackholes under consideration. Indeed, as long as we look each mass $M$ blackhole as a physical object with $\exp[A/4G]=\exp[4\pi G M^2]$ possible microstate, we will see that the probability it spontaneously radiates a particle of mass $\hbar\omega$ and becomes another object of mass $M-\hbar\omega$ equals to
\bea{}
P&&\hspace{-3mm}=\frac{e^{4\pi G(M-\hbar\omega)^2}\cdot e^{4\pi GM^2}}{(e^{4\pi GM^2}\!-\!1)\!+\!(e^{4\pi GM^2}\!-\!2)\!+\!\cdots\!+\!1\!+\!1}
\\
&&\hspace{-7mm}\!\stackrel{\hbar\omega\ll b}{=\!\!=\!\!}\!e^{-8\pi GM\hbar\omega}
\!=\!e^{-\frac{\hbar\omega}{kT_\mathrm{eff}}},~kT_\mathrm{eff}\equiv(8\pi GM)^{-1}
\nonumber
\eea
Because both the minimal blackhole and vacuum state are once degenerate, we have two $+1$ terms in the denominator. While due to averages on the possible initial states, all final state becomes equally hard/easy to be radiated into.  Considering randomness of the quantum radiation \cite{QMweinberg}, the average energy emitted in one such radiation event reads 
\beq{}
\langle E\rangle\!=\!\frac{\hbar\omega e^{-\hbar\omega/kT}\!+\!0}{e^{-\hbar\omega/kT}\!+\!1}\!=\!\frac{\hbar\omega}{e^{\hbar\omega/kT}\!+\!1},~\mathrm{for~fermions}
\eeq
\beq{}
\langle E\rangle\!=\!\frac{\sum_n n\hbar\omega e^{-n\hbar\omega/kT}}{\sum_n e^{-n\hbar\omega/kT}}\!=\!\frac{\hbar\omega}{e^{\hbar\omega/kT}\!-\!1},~\mathrm{for~bosons}
\eeq
This reproduces the particle spectrum of Hawking's radiation exactly but leaves no places for information to miss into.

Thermal features of the Hawking radiation are shown more exactly in R. Wald's 1975 paper \cite{wald1975} and the information missing puzzles are made razor-sharp in AMPS' 2012 firewall arguments \cite{fireworksAMPS2012}. However, all these works contain a fatally nonphysical assumption that, {\em the event horizon is a cut-clear, radius-definite geometric surface}. However, just as we will show in the following, a) by standard general relativity matters consisting the blackhole are not accumulating on the central point statically but are oscillating around there if their initial distributions are not so, b) by canonical quantum mechanics these consisting matters can only be at $\exp\{[k\approx1]\cdot A/4G\}$ possible discrete eigenstates all of which are of 0-measure on the central point but of nonzero and exponentially small outside the horizon, c) radial position uncertainties of the horizon surface are as large as the horizon scale itself ${\scriptstyle\Delta}x\approx GM$.

\begin{figure}[ht]
\begin{center}
\includegraphics[scale=0.7]{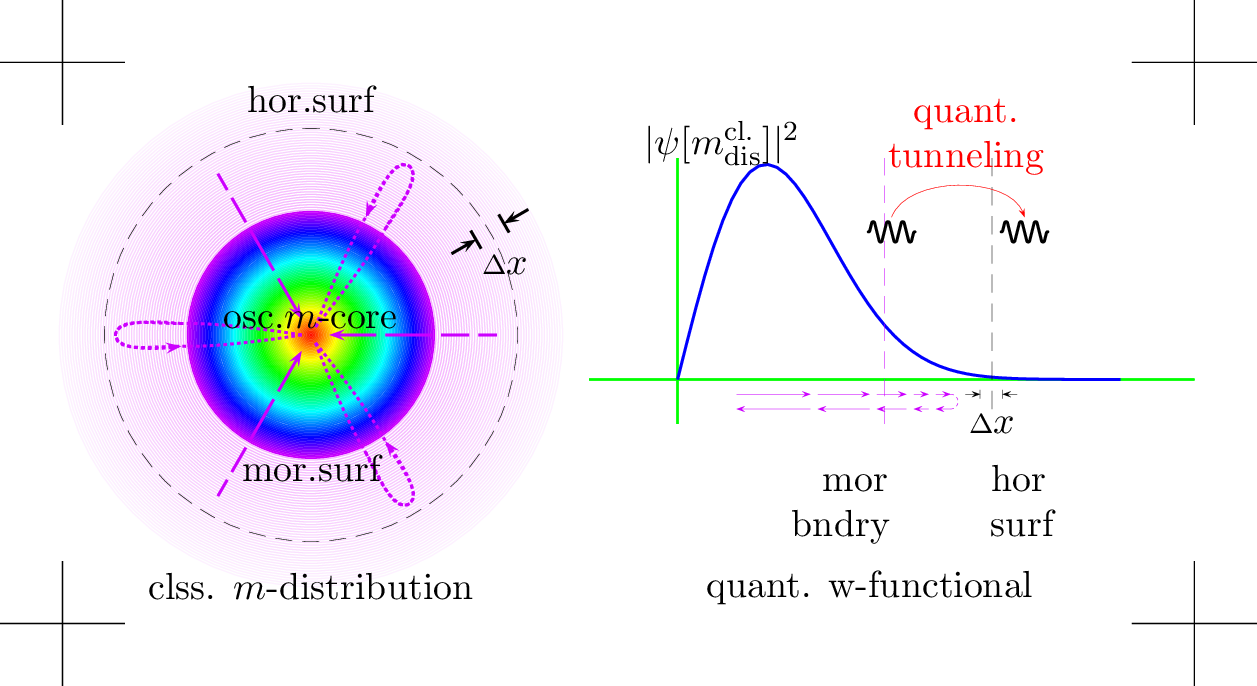}
\end{center}
\caption{The left hand side is a blackhole with oscillatory matter core and ${\scriptstyle\Delta}x$-blurred horizon surface. ``mor'' in the figure means matter occupation region. A collapsing star needs only finite coordinate time to reach radius $r_h+{\scriptstyle\Delta}x$ and will become indistinguishable from a Schwarzschild blackhole of radius $r_h$. The right hand side sketches the modular square of a typical quantum wave-functional of a matter oscillation mode. Exact such functional can be written as direct product of many factors, see eqs.\eqref{directWFproduct} and remarks therein. Each of them corresponds to the probability amplitude a mass-shell consisting the blackhole be detected at radius $r$.}
\label{figQwfunc}
\end{figure}

The last observation above is very counter-intuitive because of the usual expectation ${\scriptstyle\Delta}x\approx\lambda^M_\mathrm{deBroglie}=\hbar/M$. However, $\hbar/M$ is uncertainties of the blackhole itself (the central point position) but not the horizon-surface we are talking about. To calculate the latter, we should decompose matters consisting of the blackhole into many concentric shells and use masses of the outmost one in the de Broglie formula. From our results in \S4, eq\eqref{betaiCondition}, we know that the minimal value of such an outmost shell's mass satisfy conditions, $GMm_\mathrm{min}=\hbar~\mathrm{or}~m_\mathrm{min}=\hbar/GM$\footnote{Note that $\hbar/GM$ is also the order of hawking temperature. So such an outmost minimal-mass shell has little difference from the typical hawking particles in S-wave state.}, as results
\beq{}
{\scriptstyle\Delta}x\approx\lambda^{m_\mathrm{min}}_\mathrm{deBroglie}=\frac{\hbar}{\hbar/GM}=GM
\eeq
This uncertainty's being comparable with the horizon size itself is the key to understand our proposition in this work. It makes the horizon surface highly blurred and makes both the Hawking-Wald type calculation and the AMPS type arguments for information missing puzzle invalidate. Obviously in such case, Hilbert space of the system is not factorized into the distinctive inside and outside horizon ones. We cannot separate physics on each of them and use thermal features of the outside part to argue unitarity violation and information missing. 

The horizon surface position's uncertainty being comparable with the horizon size itself is also the key to understand a related somewhat philosophical puzzle: why we can see others around us? By general relativity, any people around us can be recognized being in a blackhole of size $H_0^{-1}$ whose center positions $H_0^{-1}$ away from us, and we ourselves are just a little away from outside the horizon, with $H_0$ here being the Hubble parameter. If the horizon is a cut-clear, radius-definite geometric surface, we will not see others because we are outside the horizon but they are inside it. This is obviously not the case. It is just the big uncertainties ${\scriptstyle\Delta}x=\hbar/m^\mathrm{min}_\mathrm{outmost~shell}=H_0^{-1}/2$ of the horizon surface's position that makes our seeing others no-barrier at all. Hawking-Wald, AMPS and many other forms of information missing puzzles all have the same logic, hence similar resolutions as this philosophical one. Obviously, if matters consisting of a blackhole are accumulating on the central point statically and form singularities there, then position uncertainties of the horizon surface will be completely the same as the blackhole central point itself and will be very small and not any resolution to this puzzle will be possible. So our resolution scheme starts from:

\section{The Schwarzschild singularity's resolving} The reason that S. Hawking neglected inner structures of the blackhole has relevance with his another  heritage, proofs of a serial of singularity theorems collaborating with R. Penrose \cite{SingularityPenrose,SingularityHawking}. According to these theorem, all matters falling across the horizon of a blackhole, including those consisting the hole itself must arrive onto the central point in finite proper time and form singularities there. Correct as it is, in both Hawking himself and the latter followers, this is over interpreted into: such singularities are static and eternal for all. 

Js as can be easily shown physically, what happens naturally when two particles colliding together or a spherical shell collapses is not a static singular point's formation, but such a point's instantaneous formation and the closely following up crumbling. For perturbation field theory and particle physicists, this is a very common knowledge because in their $S$-matrix $S=1+\mathrm{i}T$, the forward amplitudes $1$ are always the main and dominant part, while the scattering or bounding-states' formation part are always the sub-dominant and $E^{-1}$ decaying one (cross section $\sigma$ of the scattering is $E^{-2}$ decaying, $\sigma\propto|T|^2$), here $E$ is the total energy of the system. In a two-body colliding or spherical shell collapsing induced by gravitations only, $E$ is inversely proportional to the size $a$ of the region our particle or shells could be positioned by us. The idea that two particles colliding together or a shell collapsing to a point corresponds to just the case $a=0$ and $E=\infty$, in which $S=1$ and the forward scattering matters only. Obviously, in this forward process, singularities form and crumble instantaneously, no static and eternal singularities form here.

\begin{figure}[ht]
\begin{center}
\includegraphics[scale=0.8]{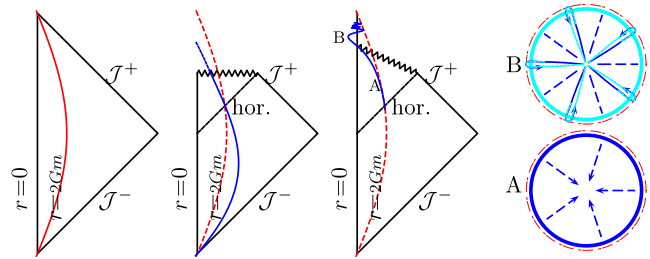}
\end{center}
\caption{The left-most part is the Penrose-Carter diagram of a Minkowski space time with a fixed size sphere. Blue curves in the left second is a freely collapsing dust shell with mass $m$ released from infinity \cite{SingularityHawking}. The popular saying is that when the shell contracts into the horizon due to its own gravitation, any game is over and nothing is knowable to the outsides. While the right most part displays a dust shell's oscillation which is released statically from its horizon. The right second is the Penrose-Carter diagram of this oscillation.}
\label{figPenrose}
\end{figure}

In general relativity, this singularity's formation and crumbling can be shown as follows. Consider a freely collapsing zero pressure spherical shell under self gravitations. Spacetimes outside such a shell is Schwarzschildian, while the inside, Minkowskian. Sizes of the shell change the same way as the radius coordinate of a freely falling test particle in a Schwarzschild background with equal masses. For observers fixed outside far away and those fixed on the central point respectively,
\beq{}
\Big\{\begin{array}{l}h\dot{t}\!=\!\gamma\!\Leftarrow\!\ddot{x}^0\!+\!\Gamma^0_{\mu\nu}\dot{x}^\mu\dot{x}^\nu\!=\!0\\h\dot{t}^2-h^{-1}\dot{r}^2=1\end{array}
\!\!,~\Big\{\begin{array}{l}^{_\prime}\!\dot{r}=\dot{r},~h=1\!-\!\frac{2Gm}{r}\\
^{\scriptscriptstyle\prime}\!\dot{t}^2-^{\scriptscriptstyle\prime}\!\!\dot{r}^2=1\end{array}
\label{classicEom}
\eeq
\beq{}
\mathrm{or}~\Big\{\begin{array}{l}\dot{r}^2=\gamma^2-h\\
\dot{t}^2=\gamma^2h^{-2}
\end{array},~
\Big\{\begin{array}{l}^{\scriptscriptstyle\prime}\!\dot{r}^2=\gamma^2-h,~h=1-\frac{2Gm}{r}\\
^{\scriptscriptstyle\prime}\!\dot{t}^2=\gamma^2-h+1
\end{array}
\label{classicEomSimp}
\eeq
where $\{t,r\}$, $\{{}^{\scriptscriptstyle\prime}\!t,{}^{\scriptscriptstyle\prime}\!r\}$ and overdot are time, spatial radial coordinates of the shell used by the two kinds of observers and derivatives with respect to proper times of the shell respectively; $\gamma^2$ is an integration constant equals to the value of $h$ on $r=r_\mathrm{rel}$ where the shell is released statically. If this release occurs outside the horizon, $\gamma$ will be real and less than $1$. While if it occurs inside the horizon, $\gamma$ will be pure imaginary and $\gamma^2<0$. Talking about release and motion inside the horizon is meaningful because we can have observers on the central point around which the spacetime is simply Minkowskian before the shell arrives onto, see e.g. \cite{stojkovic2015}. While as the shell arrives onto, its radial speed becomes that of light $\Big(\frac{d^{\scriptscriptstyle\prime\!\!}r}{d^{\scriptscriptstyle\prime}\!t}\Big)^2\xrightarrow{^{\scriptscriptstyle\prime\!\!}r\rightarrow0}1$, so it cannot be stopped there but have to go across that point and oscillate thereafter. This resolves the Schwarzschild singularity classically but avoid contradictions with Hawking and Penrose's singularity naturally. Singularities indeed form in finite proper times after the collapsing begins. However, it crumbles at the same time it forms, see the following diagram representation.

Quantum mechanically, resolving of the Schwarzschild singularity is also straightforward \cite{dfzeng2018}. We only need to multiply the shell mass on both sides of eqs.\eqref{classicEomSimp}'s first line thus translating them into Hamiltonian constraints and quantise canonically, that is, replacing $m\dot{r}\rightarrow i\hbar\frac{\partial}{\partial r}$ and introducing a wave function $\Psi(r)$ to denote the probability amplitude the shell be measured at radius $r$ and let it be functioning object of the operatorised constraint, we will get
\beq{}
[-\hbar^2\partial_r^2-m^2(\gamma^2-h)]\Psi=0,~
\gamma^2\equiv h[r_\mathrm{rel}]\in(-\infty,0)
\label{schrodingerEqOneLayer}
\eeq

It can be verified that, the finite at origin, square integrable solution to eqs.\eqref{schrodingerEqOneLayer} exists only when $\frac{Gmm\hbar^{-1}}{\sqrt{-\gamma^2+1}}=1,2,\cdots,n$. This is very similar to the simple hydrogen atoms \cite{QMweinberg} and
\beq{}
\Psi=\Psi_\beta(\hat{r})=e^{-\beta\hat{r}}\beta\hat{r} L_{q-1}^1(2\beta\hat{r})
,~\hat{r}\equiv rm\hbar^{-1}
\label{psiAllowedOneLayer}
\eeq 
\beq{}
1\leqslant\beta^2\equiv-\gamma^2+1,~\frac{2Gmm}{2\beta\hbar}\equiv q=1,2,\cdots,q_\mathrm{max}
\label{QConeLayer}
\eeq
where $L^1_{q-1}(x)$ is the first order associated Legendre polynomial. However, two big differences exists here. The first is, because the shell is constrained to be spherical and we are focusing on its radius' evolution, the question is essentially one-dimensional. As results, a factor of $r^{-1}$ is absent relative to hydrogen atoms from the probability amplitude and the wave function is definitely zero at the origin. This is nothing but the Schwarzschild singularity's resolving at quantum levels. Classically, this resolving corresponds to the fact that the shell is going across the central point at the speed of light, which is the highest value during the whole oscillation process. So the probability the shell be measured on that point takes minimal values relative to those on other places. The second difference is that, since the shell is released statically from inside the horizon thus $\gamma^2=h(r_\mathrm{rel})<0$, the value of $\beta$ is always larger than 1. This leads to that given masses, the number of allowed quantum states of the shell is finite, symbolically
\beq{}
q_\mathrm{max}=\mathrm{Floor}[Gmm\hbar^{-1}]
\label{nAllowedOneLayer}
\eeq
Even if $0<\gamma^2=h(r_\mathrm{rel})$ thus $\beta<1$, so that the shell is released from outside the horizon, as long as the release point is finitely further away from the horizon, $q_\mathrm{max}$ will be finite.

\section{The origin of Bekenstein-Hawking entropy} The spherical shell is the extremal case of general spherical symmetric objects with continuous mass distributions. For such objects, we
can decompose them into many sub-shells with quantum states of the whole system denoted by the direct product of all sub-composites. Superficially looking, this shell decomposition is arbitrary and the number of possible schemes should be non-countable. However, since the microstate of all sub-spheres are quantised similarly as eqs.\eqref{schrodingerEqOneLayer}, \eqref{psiAllowedOneLayer} and \eqref{QConeLayer}, the number of distinguishable quantum states of the system is countable and finite. Denoting the mass of all sub-spheres in a shell decomposition scheme as $\{m\}=\{m_1,m_2,\cdots,m_\ell\}$, and the mass of contents inside each shell as $\{M_1,M_2,\cdots,M_\ell\}$, with $M_1=m_1$, $M_2=m_1+m_2$, $\cdots$, $M_\ell=\sum_i^\ell m_i$, then the microstate of the system and their number will be respectively
\beq{}
\Psi=\Psi_{\{\beta\}}=\Psi_{\beta1}\otimes\Psi_{\beta2}\cdots\otimes\Psi_{\beta\ell}
\label{directWFproduct}
\eeq
\beq{}
[-\hbar^2\partial_r^2\!-\!m^2(\gamma^2_i\!-h_i)]\Psi_{\beta i}\!=\!0,\gamma^2_i\!=\!1\!-\!\frac{2G M_i}{r_\mathrm{rel}}\!<\!0
\label{schrodingerEqMultiLayer}
\eeq
\beq{}
\Psi_{\beta i}=e^{-\beta_i\hat{r}}\beta_i\hat{r}L^1_{q_i-1}(2\beta_i\hat{r}),~\hat{r}=rm\hbar^{-1}
\eeq
\beq{}
1\!\leqslant\!\beta_i\!=\!\sqrt{-\gamma_i^2\!+\!1},~\frac{2GM_im_i}{2\beta_i\hbar}\!=\!q^i\!=\!1,\!2\cdots,\!q^i_\mathrm{max}
\label{betaiCondition}
\eeq
\beq{}
W=\sum_{\{m\}} W_{\{m\}},~W_{\{m\}}=q^1_\mathrm{max}\cdot q^2_\mathrm{max}\cdots q^\ell_\mathrm{max}
\eeq
For these objects with small masses, the quantum state can be listed out by brute force. We do so in ref.\cite{dfzeng2018} and find that their number falls on the curve $W(M)=e^{k*A(M)/4G}$ very precisely. This happens to be the area law of Bekenstein-Hawking entropy formulas except for numerical factor $k\approx(8\pi)^{-1}$. 

\begin{table}
\begin{center}
\begin{tabular}{ccccc}
\hline
$\{\frac{m_i}{M_\mathrm{pl}}\}$&$\{2\}$&$\{2\m2^{\m\frac{1}{2}},2^{\m\frac{1}{2}}\}$
&$\rlap{\textcolor{red}{\rule[-1pt]{15mm}{2mm}}}{\{2\m\frac{2}{\sqrt{2}},\frac{2}{\sqrt{2}}\}}$
&$\rlap{\textcolor{red}{\rule[-1pt]{8mm}{2mm}}}{\{1,1\}}$
\\
$\{\frac{M_i^{1/2}}{M^{1/2}_\mathrm{pl}}\}$&$\{2^\frac{1}{2}\}$&$\{(2\m2^{\m\frac{1}{2}})^\frac{1}{2},2^\frac{1}{2}\}$
&$\{2\m\frac{2}{\sqrt{2}},\sqrt{2}\}$
&$\{1,1\}$
\\
${M_i^{1/2}\!m\over M_\mathrm{pl}^{3/2}}$&$\{2^\frac{1}{2}\!\!\cdot\!2\}$&$\{(2\m2^{\m\frac{1}{2}})^\frac{3}{2},1\}$
&$\{(2\m\sqrt{2})^\frac{3}{2},2\}$
&$\{1,1\}$
\\
$\{q_i\}$\!&\!$\{1\!,\!2\}$\!&\!$\{\{1\}\!,\!\{1\}\}$
\!&\!$\{\!\{0\}\!,\!\{2\}\}$
\!&\!$\{\{1\}\!,\!\{1\}\}$
\\
$\{q_\mathrm{max}\}$&$\{2\}$&$\{1,1\}$
&$\{0,1\}$
&$\{1,1\}$
\\
\hline
\end{tabular}
\end{center}
\caption{The shell decomposition and microstate quantum number of a freely collapsing dust ball with horizons in 4+1 dimensional space time. The third column decomposition is non-allowable because its most inner shell of mass $2-\sqrt{2}$ does not satisfy quantising conditions of eq.\eqref{quantiseCondition5d}. The last column should not be counted as a valid decomposition way because it cannot lead to distinguishable quantum state from the second one.}
\label{dcmp2mpl5d}
\end{table}

\begin{figure}[ht]
\begin{center}
\includegraphics[scale=0.37]{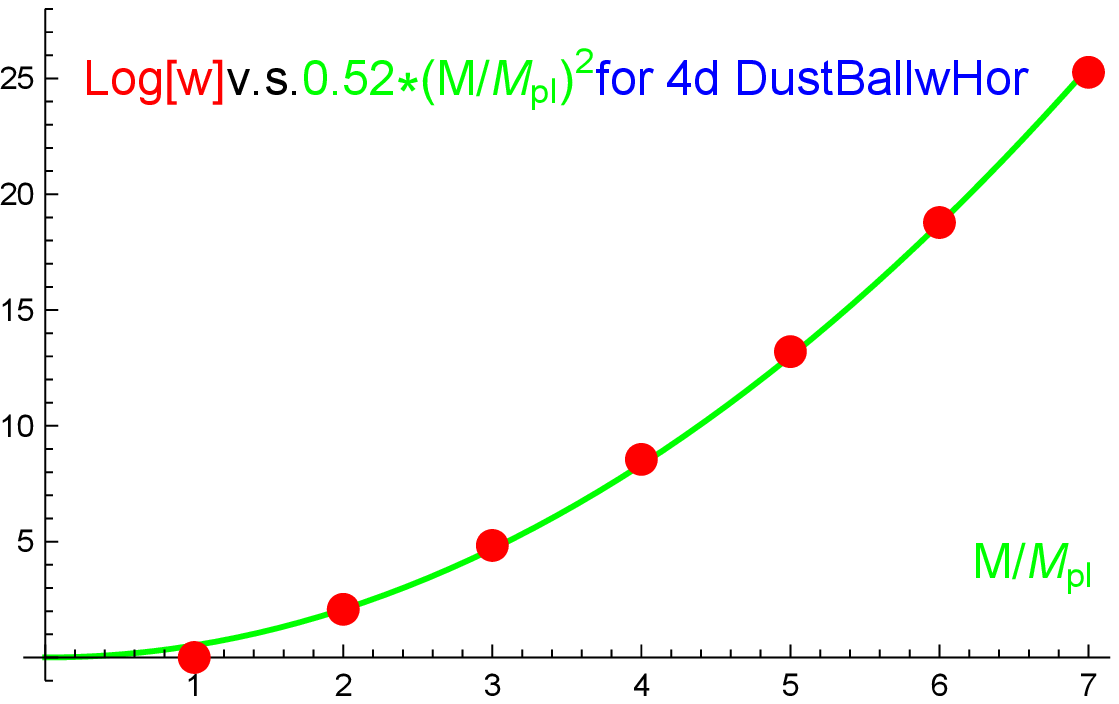}
\includegraphics[scale=0.37]{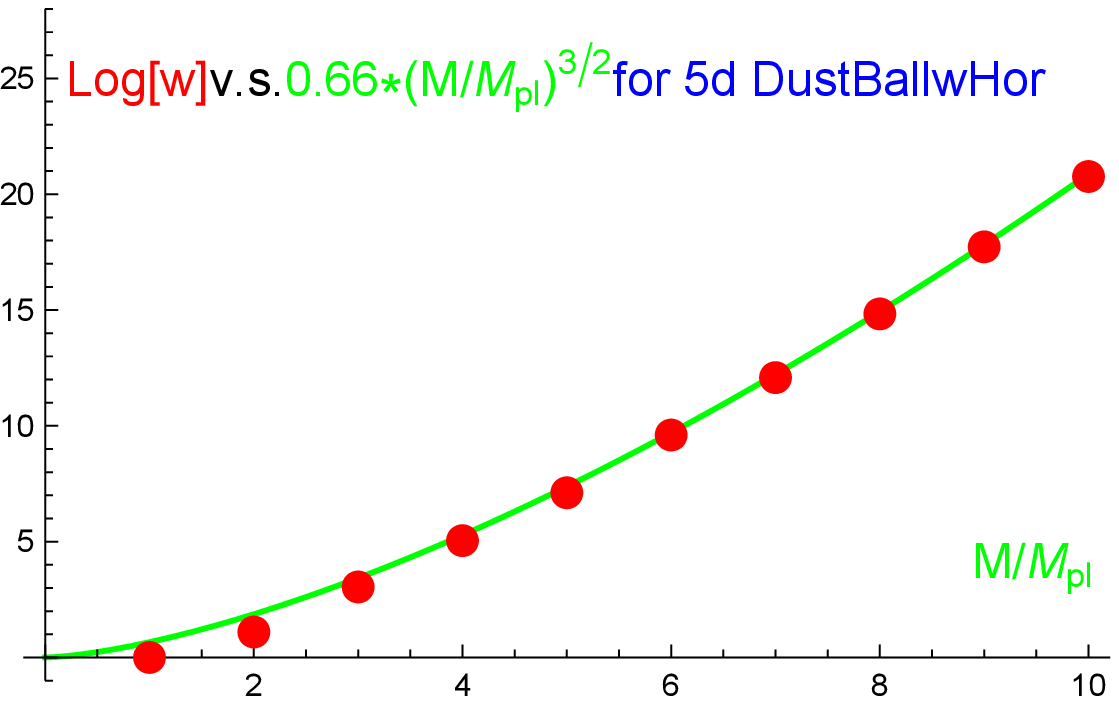}
\end{center}
\caption{The logarithmic value of microstate number v.s. area of dust ball with horizons in 4- and 5-dimensional space-time.  The scattered points are the results of this work's microstate definition and enumeration by brute force. While the continuous line is the fitting formula of $kA(M)/4G$. $A\propto M^2$ in 4d and $\propto M^\frac{3}{2}$ in 5d}
\label{figNumberBH4D5Ds}
\end{figure}

Obviously, if this quantum state definition and enumeration indeed catches the essence of blackhole microscopic degrees of freedom, we will expect that it is dimensionally universal. Just as we pointed out in ref.\cite{dfzeng2018}, simply generalising this calculation to 5 dimension encounters the very troublesome eigenvalue problem \cite{GTV2012} featured by Calogero potentials. However, if we focus on 3+$\epsilon$+1 dimensions and looke 5 as the limit of $\epsilon\rightarrow1$, then we will get proper results as expected. In this case, the function $h_i$ and parameter $\gamma_i$ in eq.\eqref{schrodingerEqMultiLayer} become
\beq{}
h_i(r)\!=\!1\!-\!\frac{2G M_i}{r^{1+\epsilon}},
\gamma^2_i\!\equiv\!h_i(r_\mathrm{rel})\!<\!0
\eeq
The corresponding wave dynamic equations can not be solved exactly. However, using Bohr-Sommerfield estimation \cite{fermionAdSdmw}, we can get the $\gamma$-quantizing condition
\beq{}
\int_0^\mathrm{horizon}\hspace{-3mm}\big\{m^2_i\big[\!-\!(\gamma^2_i\!-\!1)\!+\!\frac{2GM_i}{r^{1+\epsilon}}\big]\big\}^\frac{1}{2}dr\!=\!(q_i\!+\!\nu)\pi
\label{BohrSommerfield}
\eeq
\beq{}
\Rightarrow\frac{(GM_i)^\frac{1}{1+\epsilon}m_i}{\beta(\gamma_i)}=\!(q_i\!+\!\nu),~1\leqslant\beta_i\sim-\gamma^2_i\!+\!1
\label{quantiseCondition5d}
\eeq
with $q_i$ taking positive integer values and $0<\nu<1$. We will take $\nu=0$ for simplicity. By this quantising condition and completely the same listing method as in 4 dimensions, we can get all possible microstate quantum numbers of a given mass dust ball with horizons in 5 dimension. TABLE \ref{dcmp2mpl5d} lists all quantum numbers for the microstate of $2M_\mathrm{pl}$ mass ball as an example. Just the same as 4 dimensional case, quantum mechanically the contents of a 5 dimensional dust ball with horizon can only oscillate in a serial of finite, distinguishable eigen modes. For more higher dimensional cases, although the Bohr-Sommerfield estimation \eqref{BohrSommerfield} does not apply, it is believable that similar results should be true either.

FIG. \ref{figNumberBH4D5Ds} displays our quantitative results for the microstate number of 4 and 5-dimensional dust ball with horizons and oscillating inner contents virsus their horizon sizes directly. From the figure, we easily see that, in both 4 and 5 dimensions, the logarithmic values of these object's microstate number fit with the area law very well. Looking from outside, such dust balls with horizons are not distinguishable from the equal mass Schwarzschild blackholes. This implies that, microstates of Schwarzschild blackholes follow from the oscillation modes of its consisting matters inside the horizon. The matter/energy contents and corresponding layering structure of the pre-hole star are not destroyed on the central point, they are just over-running and going to the other side of the system at all. Two possible questions may arise here, i) why we ignored the pressure of contents of the pre-hole star? ii) why our microstate counting did not yield the Bekenstein-Hawking forumla $S=\frac{A}{4G}$ exactly but a proportional relation? On the first question, our reply is, when all known physic effects cannot resist the collapsing trends of self-gravitation, ignoring pressures of the consisting matters is reasonable. 

\begin{figure}[ht]
\begin{center}
\includegraphics[scale=0.37]{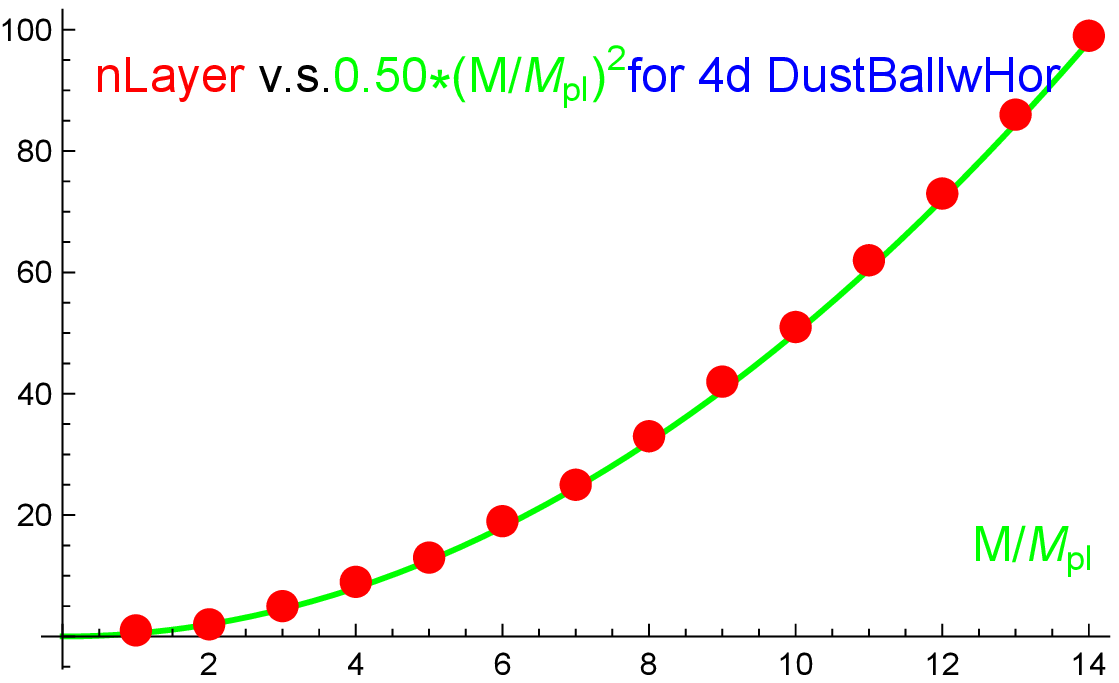}
\includegraphics[scale=0.37]{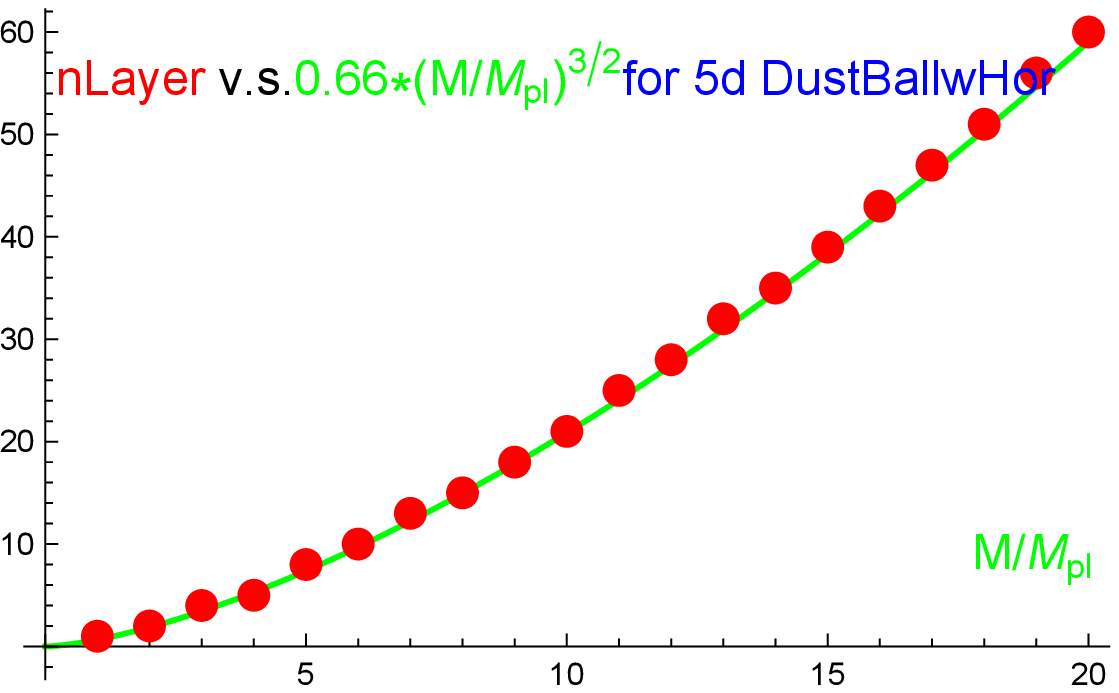}
\end{center}
\caption{The maximal number of layers a dust ball with horizon can be divided into and their masses in 4- and 5-dimensional space-time.  The scattered points is the results maximal number of layers following from the quantisation condition of \eqref{betaiCondition} and/or \eqref{quantiseCondition5d}, with all $\beta_i$ setting to 1  and counting by brute force. While the continuous line is the fitting formula of $kA(M)/4G$. $A\propto M^2$ in 4d and $\propto M^\frac{3}{2}$ in 5d}
\label{figMaxNumLayers}
\end{figure}
On the second question, we note that when a dust ball is partitioned into several concentric spherical shells, e.g. three layer of mass $\{m_1,m_2,m_3\}$, the matter contents inside each shell have their own horizon sizes, e.g. $\{r_{h1},r_{h2},r_{h3}\}\propto\{M_1,M_2,M_3\}$, $M_i\equiv\sum_{j\leqslant i}m_i$. It is classically allowable that the matter contents of layer 2 moves inside $r_{h3}$ but outside $r_{h2}$, so its integration constant $\beta_2=\sqrt{-(1-\frac{r_{h2}}{r})+1}<1$. However, the condition $1\leqslant\beta_i$ in eqs. \eqref{betaiCondition} and \eqref{quantiseCondition5d} excludes this possibility, thus underestimates the microstate number systematically. A new question arises here, will the remedy of this underestimation breaking the area law originating from its being neglected? The answer is, no. Because when this remedy of shell-crossing motion is included, all sub-shells will be equal-footing and the microstate of the system will be symmetric direct product of them and their number will be given by approximately,
\beq{}
W=\underbrace{c\cdot c\cdot \cdots\cdot c}_{\mathrm{max.~num.~layers}}=c^{\mathrm{max.num.layers}}
\eeq
The maximal number of layers such a ball can be divided into is completely determined by the quantisation condition of eqs.\eqref{betaiCondition} and/or \eqref{quantiseCondition5d}, with all $\beta_i$ being set to $1$. The results is shown in FIG.\ref{figMaxNumLayers} directly. Comparing with FIG.\ref{figNumberBH4D5Ds}, we easily see that the maximal number of layers and the logarithmic value of microstate number satisfy the same area law, even the proportionality constant. Obviously, as long as we take $c_{4d}=e^{8\pi}$ or $c_{5d}=e^{3\pi^2/4}$ similarly, will get the Bekenstein-Hawking formula for blackhole entropies exactly.

This section tells us that, contents collapsing into their own horizon are not fixed on the central point and form singularities there statically. They are classically oscillating around there periodically and quantum mechanically collapsing and expanding there in a serial of eigen-modes. Dimension universally, the number of these eigen-modes are finite and countable, and happens to be the equal mass blackhole's entropy exponentiation. This forms a very simple but intuitive microscopic interpretations for the latter.

\section{Hawking radiation, where does the information go?} Now after establishing pictures for the microstate of blackholes, we have chances to give the whole process of Hawking radiation a more explicitly unitary formulation and the question on where the information go when a blackhole evaporates a more concrete answer. According to our picture, the Hawking radiation of a blackhole is nothing but the spontaneous radiation of a quantum system with $e^{A/4G}=e^{4\pi Gm^2}$ possible initial states. Similar to all such systems \cite{JCmodel,JCmodelReview}, we can write down the hamiltonian controlling their evolution as follows
\beq{}
H=H_{\scriptscriptstyle\mathrm{BH}}+H_\mathrm{vac}+H_\mathrm{int}
\label{Hamiltonian0}
\eeq
\beq{}
=\left(\begin{matrix}b_{n}\\~\!\!&\!\!b_{n^\mm}\\~\!&\!~\!\!&\!\!\ddots
\\~&~&~&b_0\end{matrix}\right)
+\sum_k\hbar\omega_ka_k^\dagger a_k
+\sum_{|b_u\m\,b_v|}^{\hbar\omega_{\!k}=}\!\!g_{uv}b^\dagger_{uv}a_k
\label{Hamiltonian1}
\eeq
where $H_{\!_\mathrm{BH}}$, $H_\mathrm{vac}$ and $H_\mathrm{int}$ are respectively hamiltonians of the blackhole, the environment and interactions between them two. About $H_{\!_\mathrm{BH}}$, we need only know that its eigenvalues are $\{b_n,b_{n^\mm},\cdots,b_1,b_0=0\}$ and respectively $\{w\!=\!e^{4k\pi Gb_n^2}$, $w^{\m\m}=\!e^{4k\pi Gb_{n^\mm}^2}$ $\cdots,1,1\}$-times degenerating. $H_\mathrm{vac}$ is denoted by many harmonic oscillator modes, bosonic ones here for simplicity. When a mode's $\hbar\omega_k$ happens to equal to some two eigenvalue $b_n$'s difference, it will have chances to go on shell and radiated away from the blackhole. The $H_\mathrm{int}$  part functions to bring energies from the former to the latter and vice versa. Denoting the quantum state of a mass $b_\ell$, binding status $u\in\{1,2,\cdots,e^{4k\pi Gb_\ell^2}\}$, blackhole and its environment consisting of Hawking modes as $|\ell^u,n-\ell\rangle$, then\footnote{$b_v$, $b_u$ are abbreviations for $b_{\ell^{\!u}}=b_{\ell}$ and $b_{(\ell+k)^{\!v}}=b_{\ell+k}$ respectively.}
\beq{}
{\!}^{\,\hbar\omega_{\!k}\!=}_{b_{\!u}{\!\m\,}b_v\!}\!(b_{uv}^\dagger a_k)|\ell^v,n\m\ell\rangle=|(\ell+k)^u,n\m\ell\m k\rangle
\eeq
\beq{}
b_{uv}^\dagger\!=\!b_{vu}, a_{{\m}k}\!=\!a_{k}^\dagger
, g_{uv}\!=\!g_{vu}^*\!\propto\!\int\!\Psi^*_{\!\ell^v}\![r]\Psi_{\!(\ell+k)^{\!u}}\![r]dr
\label{guvDefinition}
\eeq
We will constrain ourselves to zero angular momentum radiations, so the transition $|\ell^v\!,n\m\ell\rangle\rightarrow|(\ell\!+\!k)^u\!,n\m\ell\m k\rangle$ is induced by gravitation, i.e. mass monopole interactions only. The proportionality in the definition of $g_{uv}$ and/or $g_{vu}^*$ just reflects the idea that two blackholes which have more similar microscopic wave functions could be more rapidly to transit to each other.

\begin{figure}[ht]
\begin{center}
\includegraphics[scale=0.5]{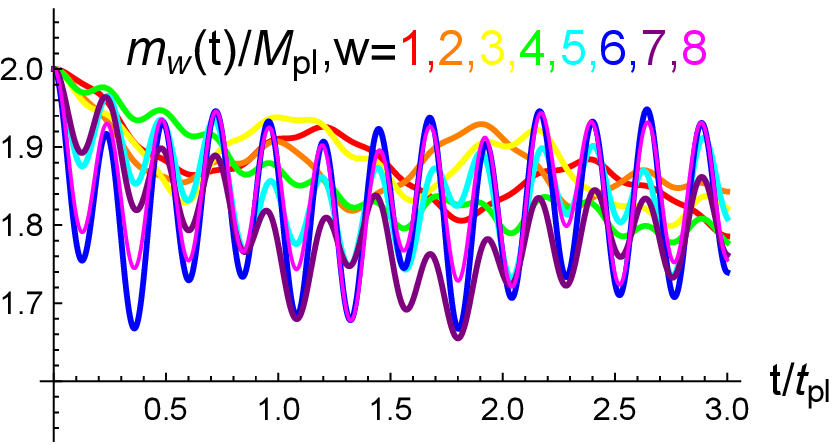}
\includegraphics[scale=0.5]{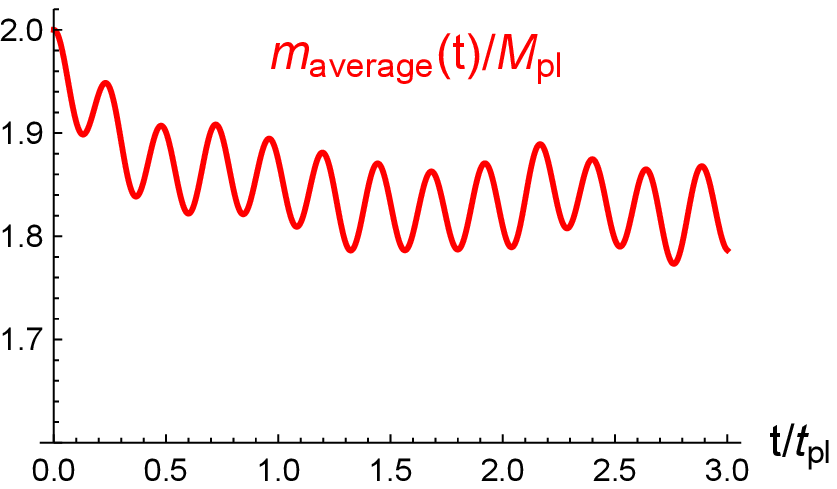}
\end{center}
\caption{The evaporating curve of 8 microstate blackholes with $2M_\mathrm{pl}$ masses as well as their average. Each curve represents a special mass variation process of the hole respectively. Since we neglect the final state Hawking particles' momentum difference, the number of final state of each micro step are highly underestimated, so the averaged evaporating curve is not monotone decreasing. }
\label{figEvapProcess}
\end{figure}

Now denoting the initial state of the system as $|n^{\!w},0\rangle$, and the latter time state as follows
\beq{}
|\psi(t)\rangle\!=\!\sum_{\ell=0}^{n}\!\sum_{v=1}^{e^{S_{\!\ell}}}e^{{\m}ib_{\ell}t}\!c_{\ell^v\!}(t)|\ell^v,n\m\ell\rangle
\eeq
we will find through Schr\"odinger equation $i\hbar{\partial}_t\psi(t)=H\psi(t)$ that
\bea{}
&&\hspace{-7mm}i\hbar\partial_tc_{\ell^v}\!(t)=(b_n\!-\!b_\ell)c_{\ell^v}\!+\!\sum_{j}^{\neq\ell}\sum_{u=1}^{e^{S_{\!j}}}g_{vu}e^{i(b_{\!\ell}-b_{\!j})t}c_{\!j^{\!u}}\!(t)
\\
&&\hspace{-7mm}c_{n^{\!w}}\!(0)\rule[2pt]{2mm}{0.5pt}1\!=\!c_{n_\m^u}(0)\!=\!c_{n_{\m\,\m}^v}(0)\!\cdots\!=\!0
\eea
This will leads to special time-dependent mass or horizon size function for each microstate blackhole
\beq{}
m_w(t)=\frac{r^w_h(t)}{2G_N}=\sum_{\ell=0}^n\sum_{v=1}^{e^{S_{\!\ell}}}b_{\ell^{\!v}}c^2_{\ell^{\!v}}(t)
\eeq
We will call this function evaporation curve of them. For a $2M_\mathrm{pl}$ mass hole, whose 8 microstates are listed out explicitly in ref.\cite{dfzeng2018}, their characteristic evaporation curve, as well as their averages are all displayed in FIG.\ref{figEvapProcess} explicitly. The average one corresponds to the variation trends of Hawking radiations. It is because we focus on too small blackholes and neglect momentum difference between the radiation particles, thus very few final states, that lead to the non monotone feature of the average curve \cite{JCmodelReview}. To get this figure numerically, we let the proportionality constant in \eqref{guvDefinition} being set to 1.

As long as we know initial status of the blackhole, i.e. the $w$ value in $n^{\!w}$, then we will be able to predict its latter time size variations exactly. On the other hand, if we precisely measure a blackhole's evaporation curve, then we will be able to infer its initial status exactly. So the information of initial blackholes goes to the evaporating hole's size variation feature, no information missing is possible here. In Hawking's calculation, it is the implicit averaging over initial hole's microstate that leads to thermal features of the radiation product and neglecting of the evaporating hole's variation process that leads to the information missing puzzle.

In some versions of the information missing puzzle, it is argued that allowing radiation processes of the blackhole to carry away its information will lead to causality violations. However, just as we analyzed in \S2, validity of this argument strongly depends on assumptions that matters consisting the blackhole are accumulating on the central point statically and the horizon is a cut-clear, position-fixed geometric surface, physics between the horizon surface and central singularity are controlled by some Schwarzschild type metrics with abnormal signatures. While in our pictures, the horizon is highly blurred ${\scriptstyle\Delta}x\approx r_h/2$ and different microstate blackholes have different radial mass profile, see Fig.\ref{figQwfunc} for references. Hawking radiation is a spontaneous process whose basic mechanism is the usual quantum tunneling. For those microstate blackhole with matters oscillating more closely around the central point, quantum tunneling of particles from their matter occupation region to the outside horizon is more harder and the corresponding blackholes will have more longer lifetime. 

\section{Comparing with string theory fuzzy ball picture} Although our microstate picture for blackholes is very similar to that of string theory fuzzy balls \cite{LuninMathur0202,mathur0502}, key differences exist between the two. Most importantly, our picture resolves the Schwarzschild singularity more thoroughly than the string theory fuzzy ball does. In ours, the static,  point-like singularity is replaced by a periodically oscillating, continuous radial mass distribution \cite{dfzeng2018}. The inner metric of blackholes has the form\footnote{We provide in ref.\cite{dfzeng2019} more explicit formulas for the inner metric of Schwarzschild blackholes consisting of dust matters.}
\bea{}
&&\hspace{-5mm}ds^2=-(h^{-1}\frac{\dot{m}^2}{m'^2}+1)d\tau^2+h^{-1}dr^2+r^2d\Omega_2^2
\\
&&\hspace{-5mm}h=1-\frac{2Gm(\tau,r)}{r},~r<r_0\equiv2Gm_\mathrm{total}
\eea
where $m(\tau,r)$ is determined by Einstein equations at the classic level and by Wheeler-de Witt equations at quantum levels. In the latter case, $m(\tau,r)$ can only take some eigen modes whose total possible number equals to the exponentiated area of the hole. While in string theory fuzzy balls, the point-like central singularity is only replaced by a dynamic string-like singularity. For example, in the NS1-P representation \cite{mathur0502}, the metric of the string whose compactification in the transverse $x_i$ space will reduce to the $4+1$ dimensional fuzzy ball reads
\bea{}
&&\hspace{-5mm}ds^2=H[-du dv\!+\!Kdv^2\!+\!2A_idx_idv]\!+\!dx_i^2\!+\!dz_a^2
\\
&&\hspace{-5mm}H=1+\frac{Q_1}{|\vec{x}-\vec{F}(t-y)|^2}
,~K=\frac{Q_1|\dot{\vec{F}}(t-y)|^2}{|\vec{x}-\vec{F}(t-y)|^2}
\\
&&\hspace{-5mm}A_i=-\frac{Q_1\dot{\vec{F}}_i(t-y)}{|\vec{x}-\vec{F}(t-y)|^2}
,~B_{uv},~B_{vi},~e^{2\phi}=\cdots
\eea
where $F_i(t-y)$ are four arbitrary functions featuring the quantum motion of the strings in transverse directions $x_i,i=1,2,3,4$. Singularities still occur along the string trajectory $\vec{x}=\vec{F}(t-y)$.

Secondly, in our microstate pictures, there are clear horizons at the classic level which divide inner and outside of blackholes. At quantum levels, the horizon is blurred by the nonzero (outside horizon) value of the wave function $|\Psi_{\beta1}(r)\otimes\Psi_{\beta2}(r)\cdots\otimes\Psi_{\beta\ell}(r)|^2$ which measures the probability of mass shells consisting the blackhole are found at radius $r$. While in string theory fuzzy ball pictures, the horizon are blurred at both classic and quantum levels. In a sense, string theory fuzzy ball provides only a ``fuzzy''- without wave function description --- picture for blackholes. Due to its lacking of exact quantum wave function, it cannot tell us exactly how degree the horizon is blurred or fuzzy. While in our pictures, exact wave function description allows us to derive out the horizon fuzziness to be ${\scriptstyle\Delta}x=r_h/2$.

Thirdly, in our microstate pictures, Schwarzschild blackhole is the most simple object whose classic inner metric and quantum wave function can be worked out explicitly. While in string theory fuzzy balls, the picture can only be established for some specially designed blackholes which correspond directly or indirectly to the NS1-P system. Schwarzschild blackholes, for its lacking of enough symmetries required by string theory method, is the hardest to exploring.

Logically, we should require proper quantum gravitation such as string theory to yield fuzzy ball picture for blackholes, instead of use string theory picture to justify rationalities of a semi-classic picture's uncovering.

\section{Conclusion} To the title question, we show that it is Hawking's neglecting or implicit averaging over the inner structure microstate of blackholes that leads to the information missing puzzle. While the negligence of microstate itself originate from an over interpretation of the singularity theorem which says that matters collapsing into their own horizon will fall on to the central point and form static singularities there for ever. Both arguments basing on perturbation theory $S$-matrix and calculations in general relativity reveal that, matters collapsing into their own horizon can go across the central point and oscillate around there. Central singularity is not a static and once forever phenomena but a periodically oscillatory and dynamic one.

We provide quantum description for matters consisting of and oscillating inside the horizon of a blackhole and show that, the radial position uncertainty of the horizon surface is of the same size as the horizon radius itself ${\scriptstyle\Delta}x\approx r_h/2$ and, the distinguishable quantum state of the system is countable and finite, with the number happens to be exponentials of the horizon area and with each microstate corresponds to a special classic oscillation modes of the contents. We show that this is the case in both four and five dimensions and argue that in extremal cases when all known physic forces cannot resist the trends of self-gravitation collapses, neglecting the pressure of pre-hole contents and looking them as dusts is reasonable. Our results provide a semi-classic fuzzy ball picture for microstate of blackholes. It is the ignorance of the high degree blurring of the horizon surface ${\scriptstyle\Delta}x\approx r_h/2$ that makes the various information missing argument difficult to resolve.

Basing on our microstate interpretation, we construct a hamiltonian thus explicitly unitary formulation for Hawking radiations. Using this formulation, we calculate the characteristic mass variation curve of 8 microstate blackholes with equal $2M_\mathrm{pl}$ mass. Each of them is distinguishable from each other. This forms a concretely answer to the question on where the information is going in Hawking radiations.

We also make comparisons between our microstate picture for blackholes and that of string theory fuzzy balls and point out their differences schematically.

\section{Prospects} The goal of looking for errors in the information missing puzzle is not to find the error itself, but to find the hints of quantum gravitations. Basing on this motivation, physicists have learned many things from this looking for, e.g. the discovering of gauge/gravity duality \cite{tHooft1993, susskind1995, madacena1997}, and entanglement/geometry equivalences \cite{RT2003, fireworksAMPS2012, EREPR}. From this aspects, we may say that no one is welcome to present too simple answers to this puzzle, although such answers are believed \cite{chapline2005,bardeen1706} to exist in many serious researches, and does not repel works starting from more fundamental theory of quantum gravitations, just as we indicated in FIG.\ref{figFramework}.

However, with the development of gravitational wave phenomenologies \cite{gw150914,gw170817a}, the question on inner structure of blackholes are no long a pure theoretical one. Gravitational waves following from binary blackholes and other compact objects provides us a possible way to measure these objects' inner structure definitely. Obviously, blackholes in our microstate picture have totally different quadrupole structure from those of conventional ones featuring by the Schwarzschild singularity and the completely empty and abnormally signatured inner horizon spacetime. As results, gravitational waves following from binary blackholes in our picture are expected to have different features from those in conventional ones \cite{gwScalarSoliton}. 

So, no matter we will become the non grata man which encounters the too simple answer to the information missing puzzle or not, it is reasonable to expect that studies on the inside horizon structure and microstate of blackholes \cite{cardoso2016,afshordi2016, cardoso2017}  will become the subject of gravitational wave era in the new decades.

\section*{Acknowledgements}
This work is supported by NSFC grant no. 11875082.
 

\end{document}